\documentclass[vecphys]{svmult}

\usepackage{graphicx}        
\usepackage[bottom]{footmisc}

\usepackage{cite}

\usepackage{youngtab}

\usepackage{slashbox}
\usepackage{latexsym}

\newcommand\be{\begin{equation}}
\newcommand\ee{\end{equation}}

\newcommand\la{\lambda}
\newcommand\g{\gamma}
\newcommand\al{\alpha}

\newcommand\lb{\bar\lambda}

\renewcommand\*{\partial}

\newcommand\punkt{\,\,.}
\newcommand\komma{\,\,,}

\newcommand\ap{{\it a priori}}
\newcommand\ie{{\it i.e.}}
\newcommand\eg{{\it e.g.}}

\newcommand\ghn{\hbox{gh}\#}
\newcommand\dimm{\hbox{dim}}

\newcommand\bsk{
}

\newcommand\lra{\longrightarrow}
\newcommand\lla{\longleftarrow}
\newcommand\larrowover[1]{\vtop{\baselineskip=0pt\lineskip=0pt
      \ialign{\hfill##\hfill\cr$\lla$\cr$#1$\cr}}}
\newcommand\rarrowover[1]{\vtop{\baselineskip=0pt\lineskip=0pt
      \ialign{\hfill##\hfill\cr$\lra$\cr$#1$\cr}}}
		
\newcommand\lalb{(\la\lb)}

\newcommand\square{\Box}

\newfam\scrfam
\batchmode\font\tenscr=rsfs10 \errorstopmode
\ifx\tenscr\nullfont
        \def\scr{\cal}
\else   
        \font\sevenscr=rsfs7
        \font\fivescr=rsfs5
        \skewchar\tenscr='177 \skewchar\sevenscr='177 \skewchar\fivescr='177
        \textfont\scrfam=\tenscr \scriptfont\scrfam=\sevenscr
        \scriptscriptfont\scrfam=\fivescr
        \def\scr{\fam\scrfam}
        
\fi

\begin{document}

\frenchspacing

\title*{Pure spinor superfields --- an overview}
\author{Martin Cederwall}
\institute{Dept. of Fundamental Physics \\
Chalmers Univ. of Technology \\
Gothenburg, Sweden \\
\texttt{martin.cederwall@chalmers.se}
}
%
%
\maketitle



\newfam\msbfam
\batchmode\font\twelvemsb=msbm10 scaled\magstep1 \errorstopmode
\ifx\twelvemsb\nullfont\def\Bbb{\bf}
        \font\fourteenbbb=cmb10 at 14pt
	\font\eightbbb=cmb10 at 8pt
	\message{Blackboard bold not available. Replacing with boldface.}
\else   \catcode`\@=11
        \font\tenmsb=msbm10 \font\sevenmsb=msbm7 \font\fivemsb=msbm5
        \textfont\msbfam=\tenmsb
        \scriptfont\msbfam=\sevenmsb \scriptscriptfont\msbfam=\fivemsb
        \def\Bbb{\relax\expandafter\Bbb@}
        \def\Bbb@#1{{\Bbb@@{#1}}}
        \def\Bbb@@#1{\fam\msbfam\relax#1}
        \catcode`\@=\active
	\font\fourteenbbb=msbm10 at 14pt
	\font\eightbbb=msbm8
\fi

\abstract{
Maximally supersymmetric theories do not allow off-shell 
superspace formulations
with traditional superfields containing a finite set of auxiliary
fields.
It has become clear that off-shell supersymmetric action formulations
of such models can be achieved by the introduction of pure spinors.   
In this talk, an overview of this formalism is given, with emphasis on
$D=10$ super-Yang--Mills theory and $D=11$ supergravity.
This a somewhat expanded version of a talk presented at 
the workshop ``Breaking of supersymmetry and ultraviolet divergences
in extended supergravity'' (BUDS), Laboratori Nazionali di Frascati,
March 25-28, 2013.
}

\section{Introduction}
\label{SecIntroduction}
The search for formalisms treating maximally supersymmetric models in
a ``covariant'' way --- covariance here taken in the sense of
manifestly exhibiting Lorentz symmetry as well as the full
supersymmetry --- has a long history.
To a large extent it has been pursued in terms of first-quantised
particle (or string) theories, with the purpose of then applying second
quantisation to obtain a covariant field theory.
Let us remind how the problem arises, first in a particle or string
theory, and then in field theory.

The Brink--Schwarz superparticle
\cite{Brink:1981nb,Casalbuoni:1976tz},
where the fermions are Lorentz spinors, exhibits a problematic mixture
of first and second class constraints, as does the Green--Schwarz
superstring
\cite{Green:1983wt}. That this must be the case is
realised already from a counting of the fermionic degrees of freedom
describing massless supermultiplets, \ie, from the 1/2-BPS property
of a massless (short) supermultiplet. There is half a spinor of first
class constraint and half a spinor of second class constraints 
\cite{Siegel:1983hh,Bengtsson:1984rw,Hori:1984ew}, 
and these can not be separated in a
Lorentz-covariant manner. The first class constraints generate the
so called $\kappa$-symmetry \cite{Siegel:1983hh}. 
Some attempts to a direct covariant treatment of the $\kappa$-symmetry
have appeared (see \eg\ refs. \cite{Bergshoeff:1990vg,Grassi:2000qs}), 
but most of the proposed
solutions to the problem have involved drastic changes of variables,
such as twistor \cite{Penrose:1972ia} methods. 

Supertwistors solve the problem of covariant quantisation of
superparticles in 3, 4, 6 and 10 dimensions
\cite{Shirafuji:1983zd,Bengtsson:1987ap,Bengtsson:1987si,Berkovits:1990yc,Cederwall:1992bi}
(see also refs. \cite{Witten:1978xx,Witten:1985nt}),
and make manifest not only super-Poincar\'e but the whole
superconformal symmetry (except, of course in $D=10$).
We mention the supertwistor track here partly since it has
similarities with our main focus of attention, pure spinors, in that
both twistors and pure spinors 
are bosonic spinors (\ie, of ``wrong'' statistics), and partly since
twistor methods (of a different flavor) have been of revived interest
later and used for amplitude calculations \cite{Witten:2003nn,Berkovits:2004hg,Berkovits:2004jj,Cachazo:2004zb,Cachazo:2004kj,Boels:2006ir,Adamo:2011pv}.  
Some works seems to point towards a deeper relation between pure
spinors and twistors \cite{Berkovits:2011gh}.
It should be mentioned that, although some
attempts have been made
\cite{Cederwall:1989su,Berkovits:1991jp}, 
twistor transform methods seem less powerful in string theory than in
particle theory, due to the massive spectrum. 

The corresponding problem is of course seen also in field theory.
There, the natural way of manifesting supersymmetry is to use
superfields, that depend not only on the bosonic coordinates $x^m$,
but also on some fermions $\theta^\mu$, that together form a
(Wess--Zumino) superspace \cite{Wess:1977fn}.
If the field theory in question is a gauge theory \cite{Brink:1976bc}, the
superfield formulation will be a gauge theory on superspace
\cite{Wess:1974jb,Ferrara:1974pu,Grimm:1977xp,Siegel:1978yi,Sohnius:1980it}, 
and if it
contains gravity
\cite{Cremmer:1977tt,Cremmer:1978km,Cremmer:1979up,Green:1982tk,Howe:1983sra}, 
it will be described as superspace geometry 
\cite{Wess:1977fn,Brink:1978iv,Dragon:1978nf,Brink:1979nt,Brink:1980az,Cremmer:1980ru,Nilsson:1981bn,Howe:1981gz,Howe:1983sra}.
In both cases, the maximally supersymmetric models (which means 16
supercharges for super-Yang--Mills theory (SYM) and 32 for
supergravity (SG)) only have on-shell formulations in superspace. This
can be stated in a couple of equivalent ways. The supersymmetry
transformations close only modulo the equations of motion. In a
component formalism, there is no set of auxiliary (non-dynamical)
fields, that can be added so that the bosonic and fermionic numbers of
fields agree off-shell and fill a representation of supersymmetry.
We will come back to the superspace formulations of 
some maximally supersymmetric models later, and examine them in more
detail, because it is precisely the traditional superspace theories
that form the basis of the pure spinor superfield formalism.

Pure spinors are interesting objects from a mathematical point of
view. The original definition by E. Cartan 
\cite{Cartan,Chevalley}
is valid in even dimensions. A Cartan pure spinor is a spinor
annihilated by half-dimensional isotropic (light-like) 
subspace. 
If the dimension is $D=2n$, then this can be expressed as
$\g^{+_i}\lambda=0$, $i=1,\ldots,n$, for a suitable choice of basis
(depending on the pure spinor $\la$). 
Here, we think of the signature of space-time as split. For euclidean
signature, take the $\g$-matrices with holomorphic indices. 
Modulo a complex scale, the pure spinor space is isomorphic to the
space of isotropic $n$-planes, which is $SO(2n)/U(n)$.
This condition can
be translated into certain bilinear conditions on the spinor. 
The first case where the pure spinor condition is non-trivial is $n=4$.
Up to $n=6$, the pure spinors form the only non-trivial orbit of the
rotation group in between the full orbit of unconstrained spinors and
the trivial orbit of 0, but for higher $n$ there are more orbits 
\cite{Igusa,Zhu,Popov},
of which the pure spinor is the most constrained.

The ``pure spinors'' we will use sometimes coincide with Cartan pure
spinors, sometimes not. The canonical example of $D=10$ SYM is an
example where they are identical. The important and defining property,
that we will give a geometric interpretation, is a bilinear identity 
$(\la\g^a\la)=0$, which in $D=10$ coincides with the constraint on a
Cartan pure spinor. Even if Cartan pure spinors are uninteresting in
$D<8$, we will encounter non-trivial ``pure spinor'' constraints
\eg\ in $D=6$ and $D=3$, essentially due to the presence of R-symmetry. 
We will also use the bilinear constraint in odd dimensions, notably
$D=11$.

We are mainly concerned with field theories, including supergravity,
and will not say much about the use of pure spinors in superstring
theory. 
From investigations of the superspace
formulation of maximally supersymmetric theories, it was early
recognised that pure spinors might have a r\^ole to play in an
off-shell formulation \cite{Nilsson:1985cm,Howe:1991mf,Howe:1991bx}.
The discovery of the precise r\^ole of pure spinors came from two
independent (but in retrospect clearly related) lines of
research. One, the covariant quantisation of the superstring, provided
a valid set of ghost variables for a covariant superstring, and
thereby also for its massless sector
\cite{Berkovits:2000fe,Berkovits:2001rb}. The other was the systematic
search for higher-derivative terms in maximally supersymmetric
theories, where revisiting the structure of the superspace constraints
revealed a cohomological structure of the deformations
\cite{Cederwall:2001bt,Cederwall:2001xk,Cederwall:2001dx}, which later
was realised to be equivalent to that of the pure spinor BRST
operator.
The latter formalism led to results on deformations of SYM 
\cite{Cederwall:2001bt,Cederwall:2001td,Cederwall:2002df}
(\eg\ the full form of the terms related to $F^4$)
as well as SG
\cite{Cederwall:2000ye,Cederwall:2000fj,Cederwall:2004cg,Howe:2003cy,Tsimpis:2008zz,Policastro:2006vt} 
models.

Pure spinor superfield models have been given for SYM 
\cite{Berkovits:2001rb,Alexandrov:2007pd,Cederwall:2001bt,Cederwall:2001td,Cederwall:2002df,Movshev:2003ib,Movshev:2009ba}  
for $D=11$ supergravity 
\cite{Cederwall:2009ez,Cederwall:2010tn} and for $D=3$ superconformal
models \cite{Cederwall:2008vd,Cederwall:2008xu,Cederwall:2009ay}. It
is quite clear that the method applies to any maximally supersymmetric
model that does not contain selfdual fields.

The wide breakthrough of the use of pure spinors in connection
with supersymmetry came with the realisation of Berkovits that they
provide a good set of variables for covariant quantisation of the
superstring 
\cite{Berkovits:2000fe,Berkovits:2005bt,Bedoya:2009np}. The formalism has been extensively
used in superstring theory, see \eg\ refs. 
\cite{Berkovits:2004px,Grassi:2004xr,Oda:2004bg,
Berkovits:2006bk,Berkovits:2006vi,Oda:2007ak,Hoogeveen:2007tu,Stahn:2007uw,Aisaka:2008vw,
Bedoya:2008yw,Mafra:2009wq,Aisaka:2009yp,Gomez:2009qd,Grassi:2009fe,Mafra:2009wi,Berkovits:2009hp,Mafra:2010gj,Park:2010xwa,Gomez:2010ad,Tonin:2010mm,Mafra:2011nv,Alencar:2011tk,Oda:2011nc,
Gomez:2012ha}. 
Applications
to supermembrane theory have also been attempted, but with less clear
results \cite{Berkovits:2002uc,Fre':2006es,Babalic:2008ga}.

This presentation takes its starting point in the traditional
superspace formulation of supersymmetric field theories. In section
\ref{SecPureSPinors} we explain
why the basics of the pure spinor superfield formalism is (almost) inherent in
the superspace formalism. We derive the BRST operator of the
linearised models. Section \ref{SecCohomology} deals with the
calculation of the field content, \ie, the BRST cohomology, which is
illustrated with some examples.
In order to formulate actions, a measure is needed, which is developed
in section \ref{SecNonMin}, based on the ``non-minimal'' variables of
Berkovits.
Section \ref{SecBV} gives the field--antifield machinery needed in order to
formulate consistent interactions. The following sections deal with gauge
fixing, necessary for quantum calculations, and with an application:
to find higher-derivative terms. 
Finally, in section \ref{SecDisc} some (hopefully) interesting open
questions and possible 
developments are mentioned.

\section{Pure spinors from superspace}
\label{SecPureSPinors}
We denote bosonic and fermionic indices in coordinate basis (``curved
indices'') by
$M,N\ldots=(m,n,\ldots;\mu,\nu,\ldots)$ 
and in Lorentz basis (``flat indices'') by 
$A,B,\ldots=(a,b,\ldots;\alpha,\beta,\ldots)$.
Wess--Zumino superspace has a torsion
\be
T_{\al\beta}{}^a=2\g^a_{\al\beta}\label{EqTorsion}
\ee 
(there might be slight formal
variations on this expression, \eg\ when there is some R-symmetry in
case of extended supersymmetry, but with a liberal interpretation 
eq. (\ref{EqTorsion})
is always true). Note that we always express components in Lorentz
indices, since fermionic directions otherwise can not be seen as spinors.
This is typically the only non-vanishing torsion component at
dimension zero (in on-shell theories), dimension here being defined so
that a bosonic derivative has dimension 1 and a fermionic $1\over2$.
In flat superspace, this statement amounts to the anticommutator
between fermionic covariant derivatives being
\be
\{D_\alpha,D_\beta\}=-T_{\alpha\beta}{}^a\*_a=-2\g^a_{\alpha\beta}\*_a\punkt
\ee
In flat space, these are the ordinary derivatives
\be
D_\alpha={\*\over\*\theta^\alpha}-(\g^a\theta)_\alpha\*_a\komma
\ee
which anticommute with the global supersymmetry generators
(superspace Killing vectors) 
\be
Q_\alpha={\*\over\*\theta^\alpha}+(\g^a\theta)_\alpha\*_a\punkt
\ee

Some special possible r\^ole of pure spinors can be seen already
here. Suppose that $\la$ is pure (in the sense mentioned in the
introduction), \ie, that
\be
(\la\g^a\la)=0\label{EqPure}\punkt
\ee
If one forms the scalar fermionic operator
\be
Q=\la^\alpha D_\alpha\komma
\label{MinimalQEq}
\ee
it becomes immediately clear from eqs. (\ref{EqTorsion}) and
(\ref{EqPure}) that  
\be
Q^2=0\punkt
\ee
It is possible to think of $Q$ as a BRST operator, and examine its
cohomology. This cohomology will be non-trivial due to the pure spinor
constraint. This will actually be the BRST operator used in the
(minimal) pure spinor formalism, and its cohomology will consist of
the physical states.

In order to understand why this happens, and why it indeed is little
more than a reformulation of the traditional superspace formalism, it
is suitable to reexamine the canonical example, $D=10$ SYM (the
procedure describes equally well its dimensional reductions)
\cite{Siegel:1978yi,Nilsson:1985cm}. For
simplicity, we will use an abelian field.

Note that we aim at going directly to the field theory, without
passing via a first-quantised superparticle model. The BRST operator
(\ref{MinimalQEq}) is not obtained as the BRST operator for some
local symmetry on the world-line of a superparticle, but postulated
more or less {\it ad hoc}. It will soon be motivated from superspace
arguments, though. Some work has been done on showing the equivalence
of the first-quantised superparticle or string 
with the formulation based on $Q$
\cite{Berkovits:2004tw,Matone:2002ft,Aisaka:2005vn}. 
We take a more pragmatic point of view --- if the correct
field theories are produced we are happy with that.

\subsection{SYM}
\label{SSSYM}
We work in $D=10$, where a chiral spinor has 16 components.
The theory starts from a gauge theory on superspace
\cite{Siegel:1978yi,Nilsson:1985cm}. 
This means
that the connection 1-form \ap\ is completely general,
\be
A=E^AA_A=E^aA_a(x,\theta)+E^\alpha A_\alpha(x,\theta)
\ee 
(where $E^A=dZ^ME_M{}^A$ is the superspace vielbein).
In order to reduce the very large number of component fields, some
constraints must be imposed. One such constraint, which goes under the
name of conventional constraint, completely expresses the superfield
$A_a$ in terms of $A_\alpha$. This is desirable, since there is
another component 1-form at level $\theta$ in $A_\alpha$, and only one
in the physical theory. The conventional constraint
is formulated in terms of the field strength, in order not to destroy
gauge symmetry, and reads (in the abelian case)
\be
\g_a^{\alpha\beta}F_{\alpha\beta}=0\punkt
\ee
Since this part of $F$ is expressed as
\be
F_{\alpha\beta}=2D_{(\alpha}A_{\beta)}+T_{\alpha\beta}{}^aA_a
\komma
\ee
the conventional constraint does exactly what it is supposed to.
Then, one is left with $A_\alpha$, the lowest-dimensional superfield,

In order to take the fields on-shell the remaining part of
$F_{\alpha\beta}$ is also set to zero. This is a selfdual 5-form. We
will not exhibit the detailed calculation here, but contend ourselves
with the well known statement the setting the dimension-0 field
strength to zero gives the equations of motion for the component
fields. These sit in the superfield at order $\theta$ (the gauge
connection) and $\theta^2$ (the fermion) (and of course also at higher
orders if they contain non-zero modes). Traditionally, to keep gauge
invariance manifest, the superfield $A_a$ is not actually
eliminated. Instead one uses the Bianchi identities for the superspace
field strength $F$, which will give the equations of motions once
$F_{\alpha\beta}=0$. This is not the path taken here. Instead we leave
$A_a$ completely aside and focus on $A_\alpha$.

We can then observe that the conditions imposed are exactly those
implied by demanding that a field $\Psi=\la^\alpha A_\alpha(x,\theta)$
is annihilated by the BRST operator $Q=\la^\alpha D_\alpha$. 
The fermionic covariant derivative acts on the superfield $A_\beta$, 
and the bilinear in
$\la$ contains only the 5-form part, due to the pure spinor
condition. In addition, gauge invariance is implemented as
$\delta_\Lambda\Psi=Q\Lambda$ (that this is true for the the bosonic
connection at level $\theta$ of course requires a small calculation), 
which makes clear that the cohomology of
$Q$ describes precisely the on-shell physical fields.
The cohomology will be examined to greater generality in the following
section.

Expanding out the $\la$-dependence of the field $\Psi$, we thus have
an infinite set of superfields,
\be
\Psi(x,\theta,\la)=\sum\limits_{n=0}^\infty
           \la^{\alpha_1}\ldots\la^{\alpha_n}
        A_{\alpha_1\ldots\alpha_n}(x,\theta)\punkt
\ee
In order for $Q=\la D$ to behave as a BRST operator, it is natural to
assign a ghost number 1 to $\la$.   
We have already mentioned that the cohomology of $Q$ at order $\la$
reproduces the gauge connection
and the fermion, subject to their linearised equations of motion 
(the remaining cohomology will be left for section
\ref{SecCohomology}). The field $\Psi$ then also carries
ghost number 1, so that the physical fields have ghost number 0.

Already at this point we see that relaxing the equations of motion
is equivalent to relaxing the condition $Q\Psi=0$. If a suitable
integration measure is found, a true off-shell formulation could be
provided by an action of the type $S\sim\int\Psi Q\Psi+\ldots$, which
will be the objective of section \ref{SecNonMin} and \ref{SecBV}.

\subsection{SG}
\label{SSSG}
What will be said in this subsection will apply to $D=11$
supergravity, and its dimensional reductions. 

A spinor in $D=11$ has 32 components. The symmetric spinor bilinears
are a 1-form, a 2-form and a 5-form. In addition to the metric field,
$D=11$ SG also contains a 3-form potential $C$ with 4-form field
strength $H=dC$ and a gravitino. The component action for the bosonic
fields,
\be
S={1\over2\kappa^2}\left(\int d^{11}x
    \left(R-{1\over48}H^2\right)+{1\over6}\int C\wedge H\wedge
    H\right)\komma
\label{SGCompAction}
\ee
contains a Chern--Simons term for $C$.
 
There are two ways of approaching the superspace construction of the
supergravity. The first one is via the actual supergeometry, examined
in
refs. \cite{Brink:1980az,Cremmer:1980ru,Howe:Weyl,Cederwall:2000ye,Cederwall:2000fj,Cederwall:2004cg}. 
Here one starts with the vielbein on superspace $E_M{}^A$ together
with a Lorentz algebra-valued connection $\Omega_M$. Just like
in the case of gauge theory, all the superfields except the one of
lowest dimension, $E_\mu{}^a$, are effectively eliminated as
independent degrees of freedom via conventional
constraints\cite{GatesStelleWest,GatesSiegel,Cederwall:2004cg}. This
is slightly more involved than in the SYM case, and we refer to
ref. \cite{Cederwall:2004cg} for a complete treatment. Essentially, by
formulating constraints on the superspace torsion,
\be
T^A=dE^A+E^B\wedge\Omega_B{}^A\komma
\ee
all connection superfields and all of the vielbein
become expressible in $E_\mu{}^a$. 
The conventional constraints reduce the possible dimension-0 torsion
$T_{\alpha\beta}^a$ (apart from
the standard part $2\g^a_{\alpha\beta}$) to the irreducible
modules 
\Yboxdim6pt
\Yvcentermath1
\be
\yng(2,1) \oplus \yng(2,1,1,1,1)\komma
\ee
where the 2 or 5 antisymmetrised indices come from the contraction of
the two spinor indices with $\gamma^{ab}$ or $\g^{abcde}$.

Like in SYM, the standard procedure for deriving the full equations of
motion is not to actually
solve for the vielbein and spin connection superfields, 
but to use torsion Bianchi identities \cite{Dragon:1978nf},
\be
DT^A=E^B\wedge R_B{}^A\komma
\ee
to obtain the equations of motion without giving up any manifest gauge
invariance.  

Suppose we now want to interpret this, at the linearised level, in
terms of pure spinors. Then we again leave all the superfield except
the lowest-dimensional one out. After converting the form index on
$E_\mu{}^a$ to a flat spinor index, we have a field
$\phi_\alpha{}^a$. It is actually only its $\gamma$-traceless part
that is not eliminated by conventional constraints.
Note that the spinor bilinears appearing above in the torsion 
$T_{\alpha\beta}{}^a$ after
conventional constraints have been used, 
the 2-form and and 5-form, are exactly those which are non-vanishing
for a pure spinor. It looks reasonable to think of the linearised
superfield $\phi_\alpha{}^a$ as appearing at order $\lambda$ in a pure spinor
superfield $\Phi^a(x,\theta,\lambda)$. The linearised equations of
motion then come from $Q\Phi^a=0$. There is only a small ingredient
missing here, namely that $\phi_\alpha{}^a$ is $\gamma$-traceless, as
are the two torsion modules. This is achieved by declaring
an equivalence relation 
\be
\Phi^a\approx\Phi^a+(\la\gamma^a\varrho)\punkt
\label{ShiftSymmetryEq}
\ee
We
call this type of equivalence relation a ``shift symmetry''
\cite{Cederwall:2008vd,Cederwall:2008xu,Cederwall:2009ez,Cederwall:2010tn,Cederwall:2011vy}, 
and we
will come back to its r\^ole in the following sections.

The other way of obtaining the linearised equations of motion is from
the 3-form $C$, which extends to a 3-form on superspace.
This method has not traditionally been used alone as a formulation of
supergravity, since the geometry (via the torsion) will enter its
Bianchi identities. Nevertheless, at the linearised level this
produces all the supergravity fields, without involving superspace
geometry; this will be made clear in section \ref{SecCohomology}. 
Without going into details about conventional constraints,
it is again the lowest-dimensional superfield that is relevant. This
is $C_{\alpha\beta\gamma}$, of dimension $-{3\over2}$, and actually
only the irreducible modules consisting of $\gamma$-traceless 2-form-
and 5-form-spinors. These modules fit perfectly in the expansion of a
scalar pure spinor superfield $\Psi(x,\theta,\lambda)$ to third order
in $\lambda$, 
\be
\Psi=\ldots+{1\over6}\lambda^\alpha\lambda^\beta\lambda^\gamma
C_{\alpha\beta\gamma}+\ldots
\ee
The linearised supergravity equations of motion come from demanding
that
\Yboxdim3pt 
\be
H_{\alpha\beta\gamma\delta}
  \bigg|{}_{\yng(2,2)\oplus\yng(2,2,1,1,1)\oplus\yng(2,2,2,2,2)}=0\komma
\ee
which is equivalent to the condition
\be
Q\Psi=0\komma
\ee
since these three irreducible modules are precisely the ones occurring
in a quadrilinear of a pure spinor.


\subsection{Summary}
\label{SubSecSuperspacesummary}
We have seen, in the two main examples of $D=10$ SYM and $D=11$ SG,
that the linearised equations of motion (and gauge symmetries)  
are reproduced precisely by considering the physical fields as part of
a pure spinor superfield with appropriate properties annihilated by
the pure spinor BRST operator $Q=\lambda D$. The price paid for this
is that interactions are (for the moment) ignored, and that only some
lowest-dimensional superfield is considered. This also means that
gauge symmetry (including diffeomorphisms and local supersymmetry in
the SG case) are not kept ``manifest'' or ``geometrical''. We will
comment more on this issue when interactions are introduced, in
section \ref{SubSecInteractionME}.

\section{Cohomology}
\label{SecCohomology}
In this section, we will take a closer look at the cohomology of the
BRST operator in the two examples of section \ref{SecPureSPinors} and
some other models. The statements about it reproducing the fields of
the models in question will be made more precise, and some interesting
structure pointing forward to a field--antifield formalism will be
pointed out.

Notice that if $\lambda$ had been unconstrained (and there was no
shift symmetry, for the case of non-scalar fields), the cohomology had
been trivial. It is the pure spinor property of $\lambda$ that gives
room for some interesting cohomology. Consider, for example, a scalar
pure spinor superfield $\Psi(x,\theta,\lambda)$, and let us for the
moment forget about the $x$-dependence. A field
$\Psi=(\lambda\gamma^a\theta)A_a$ represents cohomology: acting with
$Q$ gives
\be
Q\cdot(\lambda\gamma^a\theta)A_a
=(\lambda{\*\over\*\theta})\cdot(\lambda\gamma^a\theta)A_a
=(\lambda\gamma^a\lambda)A_a=0\komma
\ee
and it is also obvious that such a field can not be written as a
$Q$-exact expression.
In the SYM case, this cohomology is precisely the zero mode of the
gauge connection. Obviously, $\Psi$ should be taken to be fermionic.

It is clear that the algebraic properties of the pure spinor $\lambda$
play a decisive r\^ole for determining the cohomology. Indeed, as we
will see in the following subsections, a partition function for the
pure spinor contains essentially all information needed to determine
the full cohomology.

We have seen one example above of an element of the cohomology of a
scalar superfield, the zero mode of the gauge connection. We also
argued in section \ref{SSSYM} that the cohomology at order $\lambda$
precisely reproduces the fields of $D=10$ SYM, subject to the
linearised equations of motion. What is the general cohomology? One
more example is the constant field, $\Psi=c$. This is a cohomology of
ghost number 1 (given the ghost number assignment of section
\ref{SSSYM}), and given the gauge transformation of $\Psi$ it is
natural to identify it as the ghost for the 
gauge symmetry.

Both these examples concern zero mode cohomology, \ie, elements of
cohomology independent of the coordinates $x$. It turns out to be very
instructive to first consider general zero mode cohomology. Not only
is it much easier to calculate, since it is a purely algebraic
problem (the operator $Q$ reduces to
$\lambda^\alpha{\*\over\*\theta^\alpha})$), it will also give all
essential information concerning the 
full cohomology. Namely, consider a zero mode cohomology of $\Psi$ at
order $\lambda^p\theta^q$. Such a cohomology will have ghost number
$\ghn(\Psi)-p$ and dimension $\dimm(\Psi)+{1\over2}(p+q)$. 
If then $x$-dependence is
introduced, how will the corresponding cohomology behave? The only
possibility is to have some field in the same module as the zero mode,
but subject to some differential equation, an equation of motion. This
equation of motion must in turn have support in the zero mode
cohomology. This means that the zero mode cohomology can be used to
read off the possible full cohomology. If there is also a zero mode
cohomology at $\lambda^{p+1}\theta^{q+2n-1}$ (\ie, at ghost number
$\ghn(\Psi)-p-1$ and dimension 
$\dimm(\Psi)+{1\over2}(p+q)+n$), a field $\phi(x)$ in some
module determined by the zero mode cohomology at $\lambda^p\theta^q$
can be subject to a (linearised) equation of motion of the form 
$\*^n\phi=0$, given that the modules of the two zero mode cohomologies
match. The corresponding 
$x$-dependent cohomology will of course take the generic
form
\be
\Psi\sim\lambda^p(\theta^q\phi+\theta^{q+2}\*\phi+\theta^{q+4}\*^2\phi+\ldots)
\punkt
\ee

\subsection{SYM}
\label{SubSecSYM}
As mentioned, the algebraic problem of calculating the zero mode cohomology 
can be used to gain information about the full cohomology 
\cite{Berkovits:2000nn,Berkovits:2001rb,Cederwall:2001dx}.
The problem can be solved by computer methods \cite{Cederwall:2001dx}
or algebraically \cite{Berkovits:2005hy}. For the field $\Psi$ of
ghost number 1 and dimension 0, the result may be summarised in table
\ref{table.SYMPsi}, 
where the horizontal direction is the expansion in $\lambda$ (\ie,
decreasing ghost number of the component fields) and the vertical is
the expansion in $\theta$ (\ie, increasing dimension within each
superfield). The expansion of the superfields in 
$\theta$ has been shifted, so that components on the same horizontal
level have the same dimension.
The modules have been labeled by the Dynkin labels of the Lorentz
group $Spin(1,9)$.
As already discussed we see the gauge ghost at $\lambda^0$ and the
physical fields (gauge connection $A_a$ and spinor $\chi^\alpha$) 
at $\lambda^1$. In
addition there are cohomologies at $\lambda^2$ and $\lambda^3$. The
ones at $\lambda^2$ indicate, according to the discussion above, that
the physical fields are subject to equations of motion. Their
interpretation as components of the field $\Psi$ is as {\it
  antifields} $A^{*a}$ and $\chi^*_\alpha$, 
fields of ghost number $-1$ with the same dimensions as
the equations of motion. The singlet at $\lambda^3\theta^5$ is the
ghost antifield $c^*$. Its presence in cohomology in turn implies the
divergencelessness of the on-shell antifield, corresponding to
conservation of the gauge current.
This is then strong evidence that using a pure spinor to go off shell
implies introducing a Batalin--Vilkovisky field--antifield
structure. This will be formalised in detail in section \ref{SecBV}.
 
\begin{table}
\renewcommand{\arraystretch}{2}
\renewcommand{\tabcolsep}{0.05cm}
\begin{tabular}{c|ccccc}
\backslashbox{{dim~}}{{gh\#~~}}&\hspace{0.2cm}$1$&$0$&$-1$&$-2$&\hspace{0.2cm}$-3$\hspace{0.5cm}\\\hline
$0$&\hspace{0.2cm}{(00000)}&&&&\\
$\frac{1}{2}$&\hspace{0.2cm}$\bullet$&$\bullet$&               \\
           $1$&\hspace{0.2cm}$\bullet$&{(10000)}&$\bullet$&       \\
      $\frac{3}{2}$&\hspace{0.2cm}$\bullet$&$(00001)$&$\bullet$&$\bullet$\\
          $2$&$\hspace{0.2cm}\bullet$&$\bullet$&$\bullet$&$\bullet$&\hspace{0.5cm}$\bullet$\\
      $\frac{5}{2}$&\hspace{0.2cm}$\bullet$&$\bullet$&{(00010)}
				&$\bullet$&\hspace{0.5cm}$\bullet$\\
           3&$\hspace{0.2cm}\bullet$&$\bullet$&$(10000)$&$\bullet$&\hspace{0.5cm}$\bullet$\\
       $\frac{7}{2}$&\hspace{0.2cm}$\bullet$&$\bullet$&$\bullet$&$\bullet$&\hspace{0.5cm}$\bullet$\\
       $4$&\hspace{0.2cm}$\bullet$&$\bullet$&$\bullet$&$(00000)$\hspace{0.2cm}&\hspace{0.5cm}$\bullet$\\
       $\frac{9}{2}$&\hspace{0.2cm}$\bullet$&$\bullet$&$\bullet$&$\bullet$&\hspace{0.5cm}$\bullet$
\end{tabular}
\caption{The zero mode cohomology in $\Psi$ for $D=10$
  super-Yang--Mills theory. The horizontal direction represents the expansion of
  the superfield in terms of $\la$ whereas the corresponding for the
  vertical (in each row) is $\theta$ (downward). The irreducible
  representations of the component fields are listed at the positions
  which describe their ghost numbers and
  dimensions.\label{table.SYMPsi}} 
\end{table}

As argued in the beginning of the present section, there is a more
direct way of deducing the zero mode cohomology (and thereby the full
cohomology) from the partition function for a pure spinor.
Consider the expansion of a function $f(\lambda)$ in a power series
expansion in $\lambda$, just as we have done for the pure spinor
superfield. The pure spinor $\lambda$ itself is in the module
$(00001)$, and the pure spinor constraint ensures that only the module
$(0000n)$ occurs at $\lambda^n$. Therefore, the component fields in
the expansion will come in the conjugate module $R_n=(000n0)$.
A formal partition function 
\cite{Chesterman:2002ey,Chesterman:2007zz,Berkovits:2005hy}
containing all information about the
expansion is
\be
{\scr P}(t)=\bigoplus\limits_{n=0}^\infty R_n t^n
=\bigoplus\limits_{n=0}^\infty (000n0) t^n\punkt
\ee
A less refined partition function is one that only counts the
dimensions of the modules, \ie,
\be
\begin{array}{ll}
P(t)&=\sum\limits_{n=0}^\infty \dimm(000n0) t^n
=\sum\limits_{n=0}^\infty
  {1\over10}{n+7\choose7}{n+5\choose3}t^n\\
&=(1-t)^{-11}(1+t)(1+4t+t^2)\\
&=(1-t)^{-16}(1-10t^2+16t^3-16t^5+10t^6-t^8)\punkt
\end{array}
\ee
Various information can be collected here. The next to last line
indicates that the number of degrees of a pure spinor in $D=10$ is
$11$ (more on this in section \ref{SecNonMin}). 
The last line (where the factor $(1-t)^{-16}$ represents the
partition function of an unconstrained spinor) is where the zero mode
cohomology can be read off: note the agreement between the numbers in
the polynomial $1-10t^2+16t^3-16t^5+10t^6-t^8$ and the dimensions of
the modules in table \ref{table.SYMPsi}. In addition, the signs of the
monomials indicate the bosonic (plus) or fermionic (minus) character
of the cohomologies (remember that $\Psi$ is fermionic, so all signs
change).
This property is of course expressible also in the more refined
partition ${\scr P}$, which can be shown to be
\be
\begin{array}{ll}
{\scr P}(t)&=\left(\bigoplus\limits_{k=0}^\infty \vee^k(00010)t^k\right)\\
&\otimes\bigl(\bigr.(00000)\oplus(10000)(-t^2)\oplus(00001)t^3\\
&\qquad\oplus(00010)(-t^5)\oplus(10000)t^6\oplus(00000)(-t^8)
      \bigl.\bigr)\komma
\end{array}
\ee
where $\vee$ denotes the symmetric product, and the first line is the
refined partition function for an unconstrained spinor. 
This unconstrained factor can
formally be written as
$(1-t)^{-(00010)}$, see ref. \cite{CederwallPalmkvistSerre}, where the
pure spinor partition function is related to a certain Borcherds algebra.

\subsection{Supergravity}
\label{SubSecSG}
The analogous procedure can be performed for $D=11$ supergravity, and
the resulting zero mode cohomologies 
\cite{Cederwall:2001dx}
are listed in table
\ref{table.SGPsi}. This list is based on the cohomologies in a {\it
  scalar} superfield of ghost number 3 and dimension $-3$, \ie, the
field $\Psi$ of section \ref{SSSG}, based on the superspace
3-form. This field must indeed be taken as the basic field of $D=11$
supergravity, since the ``geometric field'' $\Phi^a$ does not exhibit
the gauge invariance of the $C$-field --- only the field strength $H$
appears in the torsion --- so one can not hope to reproduce the
Chern--Simons term of the action of eq. (\ref{SGCompAction}) from $\Phi^a$
alone (although the equations of motion are reproducible, one of them
being the Bianchi identity for $H$). We will not bother to write down
the detailed partition function for the $D=11$ pure spinor
\cite{Berkovits:2005hy}; the
relation to the cohomology is completely analogous to the case of SYM.

\begin{table}
\renewcommand{\arraystretch}{2}
\renewcommand{\tabcolsep}{0.05cm}
\begin{tabular}{c|ccccccccc}
\backslashbox{{dim~}}{{gh\#~~}}&\hspace{0.2cm}3&2&1&0&$-1$&$-2$&$-3$&$-4$&$\hspace{0.2cm}-5$\\\hline
$-3$&\hspace{0.2cm}{(00000)}&&&&&&&&\\
$-\frac{5}{2}$&\hspace{0.2cm}$\bullet$&$\bullet$&               &&&       &\\
           $-2$&\hspace{0.2cm}$\bullet$&{(10000)}&$\bullet$&       &&&       &\\
      $-\frac{3}{2}$&\hspace{0.2cm}$\bullet$&$\bullet$&$\bullet$&$\bullet$&&&       &\\
          $-1$&$\hspace{0.2cm}\bullet$&$\bullet$&${\raise5pt\vtop{\bsk\ialign{\hfill$#$\hfill\cr
					(01000)\cr
					(10000)\cr}}}$
			&$\bullet$&$\bullet$&&\\
      $-\frac{1}{2}$&\hspace{0.2cm}$\bullet$&$\bullet$&{(00001)}
				&$\bullet$&$\bullet$&$\bullet$&&\\
           0&$\hspace{0.2cm}\bullet$&$\bullet$&$\bullet$&${\raise9pt\vtop{\bsk\ialign{
					\hfill$#$\hfill\cr
					(00000)\cr
					(00100)\cr
					(20000)\cr}}}
				$&$\bullet$&$\bullet$&$\bullet$&&\\
       $\frac{1}{2}$&\hspace{0.2cm}$\bullet$&$\bullet$&$\bullet$&${\raise5pt\vtop{\bsk\ialign{
					\hfill$#$\hfill\cr
					(00001)\cr
					(10001)\cr}}}
				$&$\bullet$&$\bullet$&$\bullet$&$\bullet$&\\
           1&\hspace{0.2cm}$\bullet$&$\bullet$&$\bullet$&$\bullet$&$\bullet$&$\bullet$&$\bullet$&$\bullet$&\hspace{0.5cm}$\bullet$\\
      $\frac{3}{2}$&\hspace{0.2cm}$\bullet$&$\bullet$&$\bullet$&$\bullet$
				&${\raise5pt\vtop{\bsk\ialign{
					\hfill$#$\hfill\cr
					(00001)\cr
					(10001)\cr}}}$
				&$\bullet$&$\bullet$&$\bullet$&\hspace{0.5cm}$\bullet$\\
           2&\hspace{0.2cm}$\bullet$&$\bullet$&$\bullet$&$\bullet$
				&${\raise6pt\vtop{\bsk\ialign{
					\hfill$#$\hfill\cr
					(00000)\cr
					(00100)\cr
					(20000)\cr}}}
				$&$\bullet$&$\bullet$&$\bullet$&\hspace{0.5cm}$\bullet$\\
       $\frac{5}{2}$&\hspace{0.2cm}$\bullet$&$\bullet$&$\bullet$&$\bullet$&$\bullet$&{(00001)}&$\bullet$
			&$\bullet$&\hspace{0.5cm}$\bullet$\\
       	   3&\hspace{0.2cm}$\bullet$&$\bullet$&$\bullet$&$\bullet$&$\bullet$
				&${\raise5pt\vtop{\bsk\ialign{
					\hfill$#$\hfill\cr
					(01000)\cr
					(10000)\cr}}}$
				&$\bullet$&$\bullet$&\hspace{0.5cm}$\bullet$\\
       $\frac{7}{2}$&\hspace{0.2cm}$\bullet$&$\bullet$&$\bullet$&$\bullet$&$\bullet$&$\bullet$
				&$\bullet$&$\bullet$&\hspace{0.5cm}$\bullet$\\
       	   4&\hspace{0.2cm}$\bullet$&$\bullet$&$\bullet$&$\bullet$&$\bullet$&$\bullet$&{(10000)}
			&$\bullet$&\hspace{0.5cm}$\bullet$\\
       $\frac{9}{2}$&\hspace{0.2cm}$\bullet$&$\bullet$&$\bullet$&$\bullet$&$\bullet$&$\bullet$&$\bullet$
			&$\bullet$&\hspace{0.5cm}$\bullet$\\
       	   5&\hspace{0.2cm}$\bullet$&$\bullet$&$\bullet$&$\bullet$&$\bullet$&$\bullet$&$\bullet$
				&{(00000)}&\hspace{0.5cm}$\bullet$\\
       	   $\frac{11}{2}$&\hspace{0.2cm}$\bullet$&$\bullet$&$\bullet$&$\bullet$&$\bullet$&$\bullet$&$\bullet$
				&$\bullet$&\hspace{0.5cm}$\bullet$
\end{tabular}
\caption{The zero mode cohomology in $\Psi$ for $D=11$
  supergravity. \label{table.SGPsi}} 
\end{table}

The reason for $\Psi$ having ghost number $-3$ is now obvious; the
lowest cohomology represents the ghost for ghost for ghost of the the
twice reducible gauge transformations of the 3-form
field. Consequently, the ``highest'' cohomology, the corresponding
antifield, is a scalar at $\lambda^7\theta^9$. The content of table
\ref{table.SGPsi} verifies that indeed all degrees of freedom of the
supergravity are present at $\lambda^3$, 
also the gravitational ones (and even some
without local degrees of freedom, related to the Weyl invariance of
ref. \cite{Howe:Weyl}). We also note the presence of ghosts for
diffeomorphisms and local supersymmetry, appearing alongside the
ghost for tensor gauge transformations at $\lambda^2$.
As in the SYM case, the zero mode cohomology (and the partition
function) is completely symmetric with respect to exchange of fields
and antifields.

\subsection{Other models}
\label{SSOtherModels}
The method may be extended to other models. Specifically, it has been
used \cite{Cederwall:2008vd,Cederwall:2008xu,Cederwall:2009ay}
for superconformal models in $D=3$: the $N=8$
Bagger--Lambert--Gustavsson (BLG) 
\cite{Bagger:2006sk,Gustavsson:2007vu,Bagger:2007jr} and  $N=6$
Aharony--Bergman--Jafferis--Maldacena (ABJM)
\cite{Aharony:2008ug} models.
Here the Chern--Simons connection comes in one (scalar) pure spinor
superfield, and the matter multiplets in another, which, in the absence
of ghosts, comes in the same module as the scalar fields, subject to a
shift symmetry.
We refer to the papers
\cite{Cederwall:2008vd,Cederwall:2008xu,Cederwall:2009ay} for details.

We can also note that models containing selfdual fields follow part of
the pattern. Take for example the $N=(2,0)$ tensor multiplet in $D=6$.
Without exhibiting the details \cite{Cederwall:2001dx} here, we note
that the correct cohomologies for fields and ghosts are produced. When
it comes to ``antifields'', however, the pattern is broken. The
equation of motion for the tensor field is the selfduality of its
field strength, and there is no symmetry between fields and antifields
in the cohomology. Therefore, equations of motion $Q\Psi=0$ are
meaningful, but the construction of an action along the lines of
section \ref{SecBV} becomes obstructed.

\subsection{Less than maximal supersymmetry}
\label{SubSecLessSS}
The procedure sketched here is not unique for maximally supersymmetric
models, although it is there that it seems to have its highest
potential. What happens if the method is attempted for a theory with
less than maximal supersymmetry? If the pure spinors are appropriately
chosen, the traditional superspace
formulation should be reproduced also here. This is indeed the case.
If such a superspace formulation results in an {\it off-shell}
supermultiplet including auxiliary fields, 
this also happens in the pure spinor formulation. The
result, then, will be a cohomology without the antifields, since we
have argued that the presence of antifield cohomology is what puts the
physical fields on shell.

This can be illustrated by $N=(1,0)$ SYM in $D=6$
\cite{Cederwall:2008zv}. There is an $SU(2)$ R-symmetry, and with
standard assignment of Dynkin labels for $Spin(1,5)\times SU(2)$ we
let $\lambda^\alpha$ transform in the module $(001)(1)$. With the pure
spinor constraint $(\lambda\gamma^a\lambda)=0$, the only remaining
spinor bilinear is the $SU(2)$ triplet selfdual 3-form $(002)(2)$. 
Note that such a pure spinor is non-trivially constrained, unlike a
Cartan pure spinor in $D=6$, which has no R-symmetry. The
superfields in the $\lambda$ expansion of a scalar pure spinor
superfield $\Psi$ are fields $A_{\alpha_1\ldots\alpha_n}$ in
$(00n)(n)$. A direct calculation of the zero mode cohomology, or
equivalently, of the pure spinor partition function, gives at hand
that cohomology only occurs at $\lambda^0$ (the ghost) and $\lambda^1$
(the physical fields). No higher cohomologies exist, and there is no
room for equations of motion for the physical fields. The cohomology
is listed in table \ref{table.SYMSixPsi}, where it is clear that in
addition to the gauge connection and fermion field, the triplet of
auxiliary fields also appears.
\begin{table}
\renewcommand{\arraystretch}{2}
\renewcommand{\tabcolsep}{0.05cm}
\begin{tabular}{c|ccc}
\backslashbox{{dim~}}{{gh\#~~}}&\hspace{0.2cm}$1$&$0$&\hspace{0.5cm}$-1$\hspace{0.5cm}\\\hline
$0$&\hspace{0.2cm}{(000)(0)}&&\\
$\frac{1}{2}$&\hspace{0.2cm}$\bullet$&$\bullet$&               \\
           $1$&\hspace{0.2cm}$\bullet$&{(100)(0)}&\hspace{0.5cm}$\bullet$      \\
      $\frac{3}{2}$&\hspace{0.2cm}$\bullet$&$(001)(1)$&\hspace{0.5cm}$\bullet$\\
          $2$&$\hspace{0.2cm}\bullet$&$(000)(2)$&\hspace{0.5cm}$\bullet$\\
      $\frac{5}{2}$&\hspace{0.2cm}$\bullet$&$\bullet$&\hspace{0.5cm}$\bullet$
\end{tabular}
\caption{The zero mode cohomology in $\Psi$ for $D=6$ $N=(1,0)$
  super-Yang--Mills theory. \label{table.SYMSixPsi}} 
\end{table}

Since all equations of motion follow from setting the auxiliary fields
to zero, it is natural that the antifields should occur as cohomology
of a separate pure spinor superfield of dimension 2 and ghost number
$-1$ transforming as a triplet. This is indeed the case. The
antifields (or, the current multiplet) is described by a pure spinor
superfield $\Psi^{*I}$, which has a shift symmetry of the form
\be
\Psi^{*I}\approx\Psi^{*I}+(\lambda\sigma^I\rho)\punkt
\ee
The cohomology in $\Psi^*$ is the mirror of the one in $\Psi$, and
listed in table \ref{table.SYMSixPsiStar}.
\begin{table}
\renewcommand{\arraystretch}{2}
\renewcommand{\tabcolsep}{0.05cm}
\begin{tabular}{c|ccc}
\backslashbox{{dim~}}{{gh\#~~}}&\hspace{0.2cm}$-1$&$-2$&\hspace{0.5cm}$-3$\hspace{0.5cm}\\\hline
$2$&\hspace{0.2cm}{(000)(2)}&&\\
$\frac{5}{2}$&\hspace{0.2cm}$(010)(1)$&$\bullet$&               \\
           $3$&\hspace{0.2cm}{(100)(0)}&$\bullet$&\hspace{0.5cm}$\bullet$      \\
      $\frac{7}{2}$&\hspace{0.2cm}$\bullet$&$\bullet$&\hspace{0.5cm}$\bullet$\\
          $4$&$\hspace{0.2cm}\bullet$&$(000)(0)$&\hspace{0.5cm}$\bullet$\\
      $\frac{9}{2}$&\hspace{0.2cm}$\bullet$&$\bullet$&\hspace{0.5cm}$\bullet$
\end{tabular}
\caption{The zero mode cohomology in $\Psi^{*I}$ for 
the antifields of $D=6$ $N=(1,0)$
  super-Yang--Mills theory. \label{table.SYMSixPsiStar}} 
\end{table}

The condition for $\Psi$ being on-shell must be separately formulated
as another condition $s^I\Psi=0$, where $s^I$ is an
operator with ghost number $-1$ and dimension $2$, such that
$s^I\Psi$ effectively starts out with the auxiliary field
\cite{Cederwall:2008zv}.

Similar considerations could be applied to other non-maximally
supersymmetric models. It has been used to check the multiplet
structure of $D=3$, $N=8$ supergravity \cite{Cederwall:2011pu}. 
The cohomology \cite{Cederwall:SGunpub} 
of $D=10$, $N=1$ SG has also been verified to agree
with known results \cite{Nilsson:1985si,Candiello:1993di}.

\section{Pure spinor space and integration}
\label{SecNonMin}
As noted in section \ref{SSSYM}, if a reasonable (non-degenerate)
integration measure $[dZ]$ ($Z$ denoting the ordinary superspace
coordinates together with the pure spinor variables) 
can be found, an action of the form
\be
S={1\over2}\int[dZ]\Psi Q\Psi+\hbox{interactions}
\ee
will provide an off-shell formulation of the model in question, and a
solution to the problem of finding an action for maximally
supersymmetric models. In view of the discussion on cohomology of the
previous section, such an action would be a classical
Batalin--Vilkovisky (field--antifield) action (see section
\ref{SecBV}).

A measure on the pure spinor space has to fulfil a number of
requirements. First, as already noted, it has to be non-degenerate in
order that the variation of the action actually implies the equations
of motion $Q\Psi=0$. In addition, and depending on the model at hand,
there are restrictions on the dimension and ghost number of the
integration. 

For the case of $D=10$ SYM, $\Psi$ has ghost number
1 and dimension 0. Therefore $\int[dZ]$ needs to have have ghost
number $-3$, and since  
${1\over g^2}\int d^{10}x\,d^{16}\theta$ has dimension
$-4+{1\over2}\times16=4$, ``$\int[d\lambda]$'' must have dimension 4. 
Correspondingly, in $D=11$ SG, the pure spinor integration measure
must contribute ghost number $-7$ and, since 
the dimension of ${1\over\kappa^2}\int d^{11}x\, d^{32}\theta$ 
is $-2+{1\over2}\times32=14$ and that of $\Psi$ is $-3$,   
it also must give dimension $-8$.
In addition the measures should have the property that 
$\int[dZ]Q\Lambda=0$, so that BRST-trivial states have zero integral
and partial integration with respect to $Q$
is possible. 

The second thing to note is there are natural operations with
precisely these quantum numbers. If we check the highest ghost
antifield cohomology, they come at $\lambda^3\theta^5$ and
$\lambda^7\theta^9$, respectively. So, an ``integration'' that picks
out the corresponding term in the expansion of a pure spinor
superfield would have $(\ghn,\dimm)=(-3,4)$ and $(-7,8)$ respectively,
as desired. This is correct in spirit, but is still a degenerate
measure, since the expansion in $\lambda$ only contains positive
powers. Some adjustment is needed. 

The solution to this problem was
provided, for $D=10$ pure spinors, by Berkovits
\cite{Berkovits:2005bt} with the introduction of so called non-minimal
variables. By the introduction of another set of pure spinors called 
$\lb_\alpha$ and a spinor of 
fermionic variables $r_\alpha$ which is pure relative $\lb$, \ie,
fulfilling $(\lb\gamma^a r)=0$, the measure could be made
non-degenerate. Non-minimal sets of variables are quite standard when
it comes to field-antifield quantisation, but the present ones are
even more natural, even from a purely geometric point of view. Namely,
although solutions to the pure spinor constraints are complex (unless
one is in split signature), we have so far assumed that the fields
depend on $\la$ and not on $\lb$. Unless we have some kind of residue
measure, it seems more natural to integrate over the full complex
variable $(\la,\lb)$. The interpretation of the fermion $r_\alpha$ is as the
differential $d\lb_\alpha$ (with the fermionic statistics coming from
the wedge product), which obviously satisfies $(\lb\gamma^a
d\lb)=0$ \cite{Berkovits:2006vi}. When more variables are introduced,
the BRST operator must be changed accordingly in order to keep the
cohomology intact. This is done by adding a term to $Q$:
\be
Q=(\la D)+(r{\*\over\*\lb})=Q_0+(d\lb{\*\over\*\lb})=Q_0+\bar\*\komma
\ee
where $\bar\*$ is the antiholomorphic exterior derivative, the
Dolbeault operator. The cohomology is unchanged, and any cohomology will 
have a representative that is independent of $\lb$ and $d\lb$.

A field $\Psi(x,\theta;\la,\lb,d\lb)$ is then seen as an
antiholomorphic form on pure
spinor space (meaning, it can depend on both $\la$ and $\lb$, but has
only antiholomorphic indices, seen as a tensor). 
A suitable assignment of quantum numbers for $\lb$ and $d\lb$ is that
$\lb$ has ghost number $-1$ and dimension ${1\over2}$ (the opposite to
$\la$), while $d\lb$ has ghost number $0$ and dimension ${1\over2}$
(there is some irrelevant arbitrariness in the assignment, as long as it comes
out right for the BRST operator).

Suppose that the integration can be written as an integral of a form
over the pure spinor space. Since no fields contain $d\la$, the
integration measure needs to contain a top form $\Omega$ with the
maximum number of holomorphic indices.  In $D=10$, this number is 11
(see below). 
In order for partial
integration of $\bar\*$ to be allowed, this form should in addition
depend on $\la$ only, so that $\bar\*\Omega=0$.
We now try an expression for the full integral over the
non-minimal pure spinor variables,
\be
\int[d\la]X=\int\Omega\wedge X\punkt
\ee
Again counting quantum numbers (for the $D=10$ case), the $\la$ and
$\lb$ integrals cancel, while the $r$ integration 
(``removal of $d^{11}\lb$'') provides ghost number 0 and dimension
$-{11\over2}$. In order to land at the desired quantum numbers for the
integration, ghost number $-3$ and dimension $-4$, the components of 
$\Omega$ must have
ghost number $-3$ and dimension ${3\over2}$, which is accomplished by 
precisely three negative powers of $\la$, 
\be
\Omega\sim\la^{-3}d^{11}\la
\ee
(we leave it as a trivial exercise to show that the same applies to any
assignment of quantum numbers to $\lb$ and $d\lb$ that respects the
ones of $Q$, and that the assignments for $d\la$ are irrelevant).

The requirement that the holomorphic top form with $\bar\*\Omega$
exists is equivalent to the existence of a Calabi--Yau structure on
the pure spinor space, defined by $\Omega$. There is indeed a unique 
$Spin(10)$-invariant Calabi--Yau metric (up to a scale) on the pure
spinor space, 
following from the K\"ahler potential \cite{Cederwall:2011yp}
\be
K(\la,\lb)=(\la\lb)^{8/11}\punkt
\label{KahlerPot}
\ee

The pure spinor constraint may be solved in a basis where manifest $Spin(10)$
is broken to $SU(5)\times U(1)$. Then,
${\bf16}\rightarrow{\bf1}_{-5/2}\oplus{\bf10}_{-1/2}\oplus{\bf\bar5}_{3/2}$,
and a spinor is represented by a 0-form $\ell$, a 2-form $\Lambda$ and
a 4-form $M$. The pure spinor constraint reads
$\ell M-{1\over2}\Lambda\wedge\Lambda=0$, so the 11 coordinates can be
taken as $\ell$ and $\Lambda$ in a patch where $\ell\neq0$.
It is obvious that 
\be
\Omega=\ell^{-3}d\ell d^{10}\Lambda
\ee
has vanishing $U(1)$ charge, and it can be checked that it is fully
$Spin(10)$-invariant. In ref. \cite{Cederwall:2011yp}, it was checked
by explicit calculation that this is the Calabi--Yau top form
corresponding to the K\"ahler potential (\ref{KahlerPot}).
It can of course also be given a covariant form. 
The expression 
$$
\Omega\sim (\la\lb)^{-3}\lb_{\alpha_1}\lb_{\alpha_2}\lb_{\alpha_3}
{\star}T^{\alpha_1\alpha_2\alpha_3}{}_{\beta_1\ldots\beta_{11}}
d\la^{\beta_1}\wedge\ldots\wedge d\la^{\beta_{11}}
$$
is indeed independent of $\lb$ \cite{Berkovits:2004bw} (which thus can
be replaced by any constant spinor),
where the the tensor $T$ is precisely what, after dualisation of the
11 antisymmetric lower indices to 5 upper ones, defines the ghost antifield
cohomology,
\be
\begin{array}{ll}
\Psi&\sim
T_{\alpha_1\alpha_2\alpha_3,\beta_1\beta_2\beta_3\beta_4\beta_5}
\la^{\alpha_1}\la^{\alpha_2}\la^{\alpha_3}
\theta^{\beta_1}\theta^{\beta_2}\theta^{\beta_3}
     \theta^{\beta_4}\theta^{\beta_5}\\
&\sim(\la\gamma^a\theta)(\la\gamma^b\theta)(\la\gamma^c\theta)
(\theta\gamma_{abc}\theta)\punkt
\end{array}
\ee

This whole procedure may be repeated for the $D=11$ pure spinors. 
The introduction of non-minimal variables is completely analogous, 
as is the formulation of the integration in terms of a Calabi--Yau top
form.
The dimension of the pure spinor space is 23, which can be deduces
from an explicit solution similar to the one for $D=10$.
When $Spin(11)\rightarrow SU(5)\times U(1)$,
\be
{\bf32}\rightarrow{\bf1}_{-5/2}\oplus{\bf5}_{-3/2}\oplus{\bf10}_{-1/2}
\oplus{\bf\bar{10}}_{1/2}\oplus{\bf\bar5}_{3/2}\oplus{\bf1}_{5/2}\punkt
\ee
A spinor is thus parametrised by an arbitrary form. If we write it as
\be
\la=\ell\oplus\bigoplus\limits_{p=1}^5\Lambda_p
\ee
($\ell$ being the 0-form, and the subscript $p$ denoting form degree),
the solution to the pure spinor constraint is
\be
\begin{array}{ll}
\Lambda_3&=\ell^{-1}\Lambda_1\wedge\Lambda_2+\Sigma\komma\\
\Lambda_4&=\ell^{-1}(-\Lambda_1\wedge\Lambda_3
     +{1\over2}\Lambda_2\wedge\Lambda_2)\komma\\
\Lambda_5&=\ell^{-2}\Lambda_2\wedge\Lambda_3
     -{1\over2}\Lambda_1\wedge\Lambda_2\wedge\Lambda_2\komma
\end{array}
\ee
where $\Sigma$ is a 3-from satisfying
\be
\imath_v\Sigma\wedge\Sigma=0
\ee
for all vectors $v$, \ie, $\epsilon^{jklmn}\Sigma_{ijk}\Sigma_{lmn}=0$
\cite{Berkovits:2005hy,Cederwall:2012es}.

An important difference compared to the $D=10$ pure spinors is that
there is a singular locus 
away from the origin, where the 3-form $\Sigma$ vanishes. It is
straightforward to see that then
$(\la\gamma^{ab}\la)=0$. This
is the space of $D=12$ Cartan pure spinors, a
16-dimensional space. The degrees of freedom contained in $\Sigma$
consists, modulo a scale, of the Grassmannian 
$Gr(2,5)={SU(5)\over S(U(3)\times U(2))}$ of 2-planes in
5-dimensions. So the appearance of $\Sigma$ provides 14 more real, or
7 complex dimensions, to make a total of 23.
A similar parametrisation of the solution of the constraint on
$\Sigma$ in terms of modules of $su(3)\oplus su(2)\oplus u(1)$, with
$s$ being the singlet, gives at hand that the the measure, \ie,
the holomorphic top form carries the factor $\ell^{-5}s^{-2}$
\cite{Cederwall:2012es}, and
here is the ghost number $-7$ as announced.  
Again, the measure can be cast in a Lorentz-covariant form, but we
will not go into the details
(see refs. \cite{Berkovits:2002uc,Anguelova:2004pg,Cederwall:2009ez}).
The above reflects the fact that the top
cohomology at $\la^7\theta^9$ contains 2 powers of
$(\la\gamma^{(2)}\la)$.
The corresponding K\"ahler potential and metric have not been
explicitly constructed, but this should be straightforward.

We finally want to say a few words about integration and
regularisation \cite{Berkovits:2005bt}. 
It was mentioned that the cohomology, also after the
introduction of $(\lb,d\lb)$, has representatives that are independent
of these variables. In other words, they are holomorphic functions
(0-forms).How can integrals of (products of) such functions give a
non-vanishing result? One will always obtain 0, due the
undersaturation of the form degree (the fermionic variables). On the
other hand, the polynomial behaviour of the cohomologies at infinity
gives $\infty$, if radial integration is performed first. The
integrals are ill-defined, of the form $0\times\infty$. This can been
remedied in two (equivalent) ways. Either we note that the
representatives in the minimal variables are a bad choice, and change
them into some BRST-equivalent representatives that give well-defined
integrals, or we use a BRST-invariant regularisation of the
measure. The same type of regulator, an expression of the form
$e^{-t\{Q,\chi\}}\komma$
works in both cases. A standard choice for $\chi$ is
$\chi=\theta^\alpha\lb_\alpha$, giving a regulator
\be
e^{-t((\la\lb)+(\theta d\lb))}\punkt
\ee
If such a regulated measure (with $t>0$) 
is used with the minimal representatives,
we see that it regulates the bosonic integrals at infinity. At the
same time 11 ($D=10$) or 23 ($D=11$) $d\lb$'s are needed to saturate
the form degree (fermionic integral), and the corresponding term in
the expansion of the exponential carries $11$ ($23$) $\theta$'s. In
order to saturate the $\theta$ integration, another 5 (9) are needed,
and we see that this agrees with picking out the top cohomology, as
was the first, too na\"ive, candidate for integration.
It is thus no coincidence that the number of $\theta$'s in the top
cohomology agrees with the number of independent constraints on a pure
spinor.

The regulated integrals will of course be independent of the parameter
$t$. This looks much like localisation --- taking $t$ to be very big
localises the integral close to the origin. The dependence on the pure
spinor variables is indeed ``topological'', in the sense that they do
not provide new functional dependence, only a finite spectrum. We have
not seen any good way of making use of localisation. The origin is
not a regular point in pure spinor space, rather a boundary
\cite{Nekrasov:2005wg}.

\begin{figure}
\centering
\includegraphics[height=5cm]{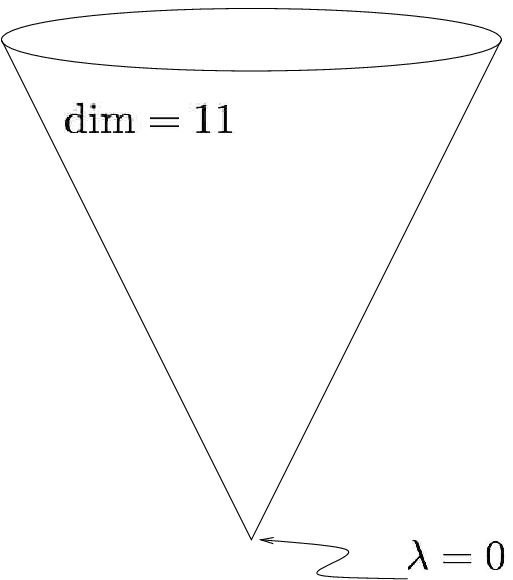}
\hskip5mm
\includegraphics[height=5cm]{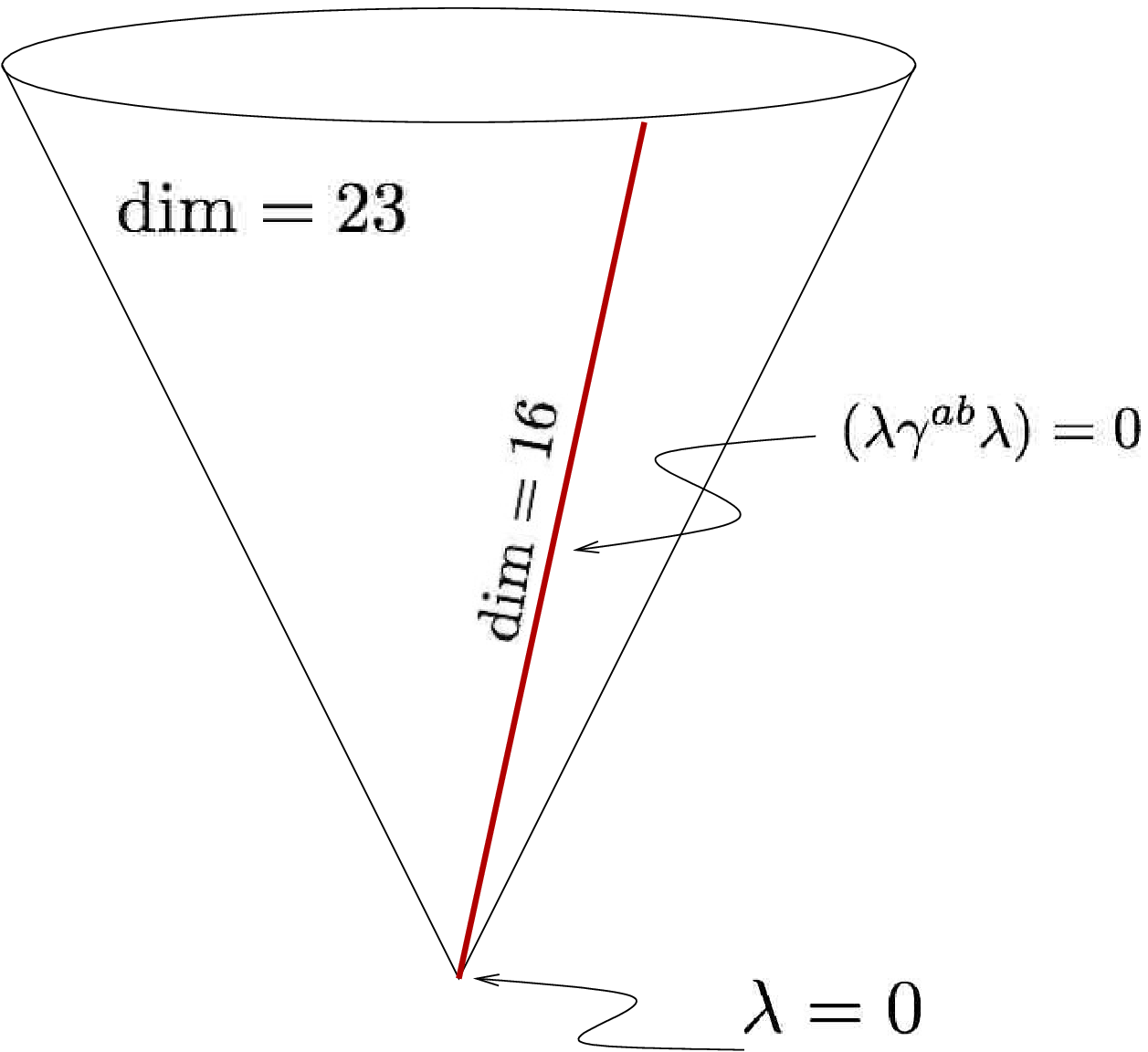}
\caption{A rough sketch of the $D=10$ and $D=11$ pure spinor spaces,
  with their respective singular subspaces marked.}
\label{fig:1}       
\end{figure}

\section{Batalin--Vilkovisky formalism and actions}
\label{SecBV}
We have seen in section 3 that the content of the pure spinor
superfields is not only the physical fields, but also a full set of
ghosts and antifields (at least for maximal supersymmetry). This
indicates that the proper framework for introducing interactions
(so far, everything has been at a linearised level) is the
Batalin--Vilkovisky formalism
\cite{Batalin:1981jr,Henneaux:1992ig,Fuster:2005eg}. 

\subsection{Field-antifield structure}
\label{SubSecFieldAntifield}
The Batalin--Vilkovisky (BV) formalism can be thought of in several
ways. It seems to have originated as an attempt to find something
similar to a Hamiltonian formalism, without breaking manifest Lorentz
symmetry, in that sense uniting the advantages of the Lagrange and
Hamilton methods. Another way of viewing it is that it naturally lifts
the BRST method to possible include nonlinear terms and
transformations, \ie, interactions. It should be noted that some
textbooks (\eg\ ref. \cite{Weinberg:1996kr}) introduce the BV
formalism in connection with gauge fixing, which tends to somewhat
obscure the simplicity. What we will do here is classical BV field theory,
although we will discuss gauge fixing in section \ref{SecGauge}.

In the BV framework, a ghost field is introduced for each gauge
symmetry (and reducibility) and each of the fields $\phi^I$ (by which is meant
physical fields as well as ghosts) is supplemented by its antifield 
$\phi^*_I$ with opposite statistics and a ghost number assignment fulfilling
$\ghn(\phi)+\ghn(\phi*)=-1$.
A fermionic bracket, the so called antibracket, between functions of
fields and antifields is introduced as
\be
(A,B)=\int d^Dx\left(
A{\larrowover\delta\over\delta\phi^I(x)}
        {\rarrowover\delta\over\delta\phi^*_I(x)}B
-A{\larrowover\delta\over\delta\phi^*_I(x)}
        {\rarrowover\delta\over\delta\phi^I(x)}B
\right)\punkt
\ee
The (classical) BV action is defined as a solution to the master
equation
\be
(S,S)=0\komma
\label{MasterEq}
\ee
which reduces to the action for the physical fields when ghosts and
antifields are removed.
The action itself generates gauge transformations via the antibracket
(in a generalised
sense, where \eg\ antifields are transformed by the
equations of motion for the physical fields), so the master equation
(\ref{MasterEq}) can be seen as the invariance of the action itself.

In the situation at hand, with the pure spinor superfields for
maximally supersymmetric theories, we have seen that the cohomology
describes both fields and antifields, so a split in the two sets looks
problematic. In addition, it is of course necessary to define the
antibracket off shell, so that also components outside cohomology
takes part. The field--antifield symmetry of the cohomology makes it
natural to think of a field $\Psi$ as self-conjugate with respect to
the antibracket, and define it as \cite{Cederwall:2009ez}
\be
(A,B)=\int A{\larrowover\delta\over\delta\Psi(Z)}[dZ]
        {\rarrowover\delta\over\delta\Psi(Z)}B
\punkt
\ee
It is straightforward to show that this antibracket (in all cases we
have considered) carries the correct quantum numbers, and that a free action
of the form
\be
S_2={1\over2}\int[dZ]\Psi Q\Psi
\ee
indeed generates gauge transformations. At this non-interacting level,
the master equation is equivalent to the nilpotency of the BRST
operator. 
Actions of this form thus describes both SYM and SG at linearised
order.

\subsection{Interactions from the master equation}
\label{SubSecInteractionME}
We now have at our disposal all ingredients necessary to introduce
interactions in a consistent way. The guiding principle is the master
equation (\ref{MasterEq}).

\subsubsection{SYM}
\label{SubsubSYM}
The SYM case is easy. The linearised action has the form of an abelian
Chern--Simons action, and since $\Psi$ and $Q$ carry the same quantum
numbers a $\Psi^3$ term can be added, turning the full action into
Chern--Simons form,
\be
S=\int[dZ]\,\hbox{tr}\left({1\over2}\Psi Q\Psi+{1\over3}\Psi^3\right)\punkt
\ee
This leads to equations of motion
\be
Q\Psi+\Psi^2=0\komma
\ee
which could of course equally well be directly deduced from the 
superspace formalism, where its restriction to the ghost number zero
fields reads
$\la^\alpha\la^\beta F_{\alpha\beta}=0$.

A notable feature is that although the component action contains
4-point couplings, such terms are not present in the manifestly
supersymmetric pure spinor superfield action. Instead they are
reproduced when the equations of motion are solved sequentially in the
$\theta$ expansion of the superfields $A_\alpha$.
Such simplifications are typical. We mentioned them in passing for the
3-dimensional conformal models of section \ref{SSOtherModels}, and
similar simplifications turn out to happen also for supergravity.

\subsubsection{SG}
\label{SubsubSG}
The interactions of $D=11$ supergravity
\cite{Cederwall:2009ez,Cederwall:2010tn} are more subtle.
Remember that $Q$ has ghost number 1 and dimension 0, while $\Psi$ has
ghost number 3 and dimension $-3$. 
The first step will be to construct a 3-point coupling. How can it be
formed, given that the integrand in the action must have ghost number
7 and dimension $-6$?

Here, the geometric field $\Phi^a$ comes into play. We remind that it has ghost
number 1 and dimension $-1$. It contains the field strength $H$ but
not the potential $C$. Guided by the form of the Chern--Simons term 
$C\wedge H\wedge H$, is it possible that something like $\Psi\Phi^2$
may work? Such a combination has ghost number 5 and dimension $-5$. If
it is supplemented by two powers of $\la$, the quantum numbers are the
correct ones. A hypothetical 3-point coupling is then
\be
S_3\sim\int[dZ](\la\gamma_{ab}\la)\Psi\Phi^a\Phi^b\punkt
\label{TentThreePoint}
\ee
Apart from the matching of quantum numbers, the factor
$(\la\gamma_{ab}\la)$ has two other r\^oles: the antisymmetry in
$[ab]$ makes it possible to contract the indices on the (fermionic)
$\Phi$ fields; and it ensures the invariance under the shift symmetry
of eq. (\ref{ShiftSymmetryEq}), thanks to the Fierz identity 
$(\gamma^b\la)_\alpha(\la\gamma_{ab}\la)=0$, satisfied by a pure
spinor $\la$ (but not by an unconstrained one). 

This is of course not the final answer for
the 3-point coupling. We have argued that $\Psi$ is the fundamental
field, but eq. (\ref{TentThreePoint}) is meaningless until we declare
how $\Phi^a$ is formed from $\Psi$. Let us assume that there is some
operator $R^a$ of ghost number $-2$ and dimension 2 
(defined modulo shift symmetry) such that 
\be
\Phi^a=R^a\Psi\punkt
\ee
Then the master equation, stating the consistency of the tentative
3-point coupling, demands that $[Q,R^a]=0$ (again modulo shift
symmetry).
Such an operator was constructed in ref. \cite{Cederwall:2009ez}, and
it takes the form
\be
R^a=\eta^{-1}(\lb\gamma^{ab}\lb)\*_b+\ldots\komma
\ee
where $\eta=(\la\gamma^{ab}\la)(\lb\gamma_{ab}\lb)$ is the scalar
invariant vanishing on the the codimension-7 subspace of
12-dimensional pure spinors, and where the ellipsis denotes terms with
$d\lb$ and $d\lb^2$.

This means that we have a consistent 3-point interaction. It is
clearly also non-trivial, and since already the 3-point coupling for
gravity is cohomologically unique \cite{Boulanger:2000rq}, 
it must be the full 3-point
coupling of $D=11$ SG in Minkowski space. A concrete check on component
field couplings would nevertheless be encouraging. In refs. 
\cite{Cederwall:2009ez,Cederwall:2010tn}, it has been verified that
the Chern--Simons term is correctly reproduced, and that the ghost
couplings corresponding to the diffeomorphism algebra are the right
ones.

Surprisingly, the 3-point interactions provide almost the full
answer. When checking the master equation to higher order in $\Psi$, a
very simple 4-point coupling arises, containing a simple nilpotent
operator $T$. The properties of this operator ensures that the master
equation is satisfied to all orders, and the full action for $D=11$ SG
is
\be 
S=\int[dZ]\left[{1\over2}\Psi Q\Psi
+{1\over6}(\la\g_{ab}\la)\left(1-{3\over2}T\Psi\right)
\Psi R^a\Psi R^b\Psi
\right]\punkt
\ee
We refer to
ref. \cite{Cederwall:2010tn} for the details.

Strikingly enough, the full action for $D=11$ supergravity becomes
polynomial. The 4-point coupling may even be removed by a field
redefinition (at the price of having a redefined field which is not
canonical with respect to the antibracket, and has a less standard
kinetic term).
However, it should be said that geometry is somewhat obscured. By
basing the formulation on the lowest-dimensional part of the
superspace fields, and treating the fields as deformation of the flat
background, geometry is not manifest. Still, the appearance of all
ghosts, including the ones for diffeomorphisms and local
supersymmetry, in the cohomology, together with the master equation,
ensures full gauge invariance, although in a form that is not easily
recognisable as geometric. 
Therefore it may be interesting to try to ``rebuild'' a geometric
picture based on the present formalism. We do not have any concrete
ideas about how this may be done, but it might involve further
variables, reintroducing the superfields that were discarded (the
higher-dimensional parts of the super-vielbein).
Formally, an analogue statement is true for the SYM action, but the simple
Chern--Simons form there makes gauge invariance (almost) manifest.
In close connection with this, it is not clear how to best find
solutions to the equations of motion. It is not known even how to
embed simple, purely gravitational, solutions like the Schwarzschild
geometry into the superfield $\Psi$. For perturbation theory around
flat space, on the other hand, the formulation is ideal, both for
keeping control over the symmetries and for having a very limited
number of couplings, and it has
been used for amplitude calculations
\cite{Cederwall:2012es,Karlsson2013}.

\subsubsection{Other models}
\label{SubsubOther}

Actions, along the lines drawn up here,
can also be constructed for the BLG and ABJM models described briefly
in section \ref{SSOtherModels}. 
Since the fields describing the scalar multiplets are non-scalar,
their kinetic terms contain extra $\la$'s ensuring shift symmetry.
The interactions consist essentially of a
minimal coupling to the Chern--Simons field, replacing and reproducing
the higher order interactions among the component fields (\eg\ a sixth
order potential in the scalars).
We again refer to refs.
\cite{Cederwall:2008vd,Cederwall:2008xu,Cederwall:2009ay} for details.

In principle, actions could be formed also for models with less
supersymmetry. Then we know from the discussion in section
\ref{SubSecLessSS} that separate pure spinor superfields must be
introduced for the fields and the antifields.  
The full formalism for lower supersymmetry has not been developed. In
ref. \cite{Cederwall:2008zv} minimal $D=6$ SYM was treated, but only
at the level of equations of motion, and in a minimal pure spinor 
formalism. Especially issues concerning gauge fixing may turn out to
be easier in such models (see section \ref{SecGauge}).
In particular, $D=10$, $N=1$ supergravity and its dimensional
reductions may be interesting, \eg\ 
concerning the investigation of possible counterterms.

\section{Higher derivative terms and Born--Infeld theory}
\label{SecBI}
As an example of an application of our formalism, we will briefly
describe the construction of a higher-derivative term. Even though the
example is specific --- the $F^4$ deformation of $D=10$, $N=1$
SYM, it may be applied to any supersymmetric
deformation of a maximally supersymmetric model with a pure spinor
action. As we will see, the drastic simplifications of interaction
terms persist also here, and although an $F^4$ deformation in
component language will come together with an infinite number of terms
of arbitrarily high order in derivatives, a single quartic term turns out to
contain the full deformation in the pure spinor superfield
language for the abelian model. 
We conjecture that it describes Born--Infeld theory.

The question addressed here was actually
one starting point for the development of the present formalism
\cite{Cederwall:2001bt,Cederwall:2001xk,Cederwall:2001dx,Cederwall:2001td,Cederwall:2002df}. The
work described in this section is based on
ref. \cite{Cederwall:2011vy}.

Precisely as for any interaction term, the guide to consistent
deformation is the master equation. What is needed is some Ansatz for
the form of the interactions. 
In refs. \cite{Cederwall:2001bt,Cederwall:2001td}, it was observed
that the 5-form part of $F_{\alpha\beta}=0$ must be changed in order
to deform the theory. It was also noted that the appropriate 
$\alpha'^2F^4$ terms for SYM were generated by 
\be
F_{\alpha\beta}^A\sim \alpha'^2 t^A{}_{BCD}
   (\gamma^a\chi^B)_\alpha(\gamma^b\chi^C)_\beta F^D_{ab}\komma
\label{FFourDef}
\ee
where $t$ is a symmetric invariant tensor, and $\chi$ and $F$ denote
the superfields with the corresponding component fields as lowest
components. 
We will from now on drop the explicit factor $\alpha'^2$. 
This was then used in ref. \cite{Cederwall:2001td} in order to derive
for the first time the complete deformation at this order, including
all fermion couplings.

We need some systematics for lifting expressions like eq. (\ref{FFourDef})
to full pure spinor superfield expressions, containing not only fields
of definite ghost number. 
The method introduced in ref. \cite{Cederwall:2011vy} was to form
``physical operators'', solving this problem. Take for example the
physical fermion. We would like to find an operator $\hat\chi^\alpha$
that, roughly 
speaking, strips the pure spinor superfield $\Psi$ of one power of
$\la$ and two powers of $\theta$ and forms a pure spinor superfield
that ``starts'' with $\chi^\alpha$, and similarly for other component
fields. These operators were systematically constructed in the
non-minimal formalism. For example, the operator $\hat\chi^\alpha$
takes the form
\be
\hat\chi^\alpha={1\over2}(\la\lb)^{-1}(\gamma^a\lb)^\alpha\*_a+\ldots
\komma
\ee
with the ellipsis denoting terms with more singular behaviour in 
$(\la\lb)$ and with one or two powers of $d\lb$.
The physical operators turn out to satisfy a number of interesting
algebraic and differential relations (among them, a somewhat
surprising relation to the $b$ operator of section \ref{SecGauge}).

We found that a quartic term in the action
\be
S_4={1\over4}\int[dZ]\Psi
(\la\g^a\hat\chi)\Psi(\la\g^b\hat\chi)\Psi\hat F_{ab}\Psi
\ee
solves the master equation in the Maxwell case, not only to this order
but to all orders, and conjectured that it describe supersymmetric
Born--Infeld theory. In the non-abelian case, the same term, dressed
up with a four-index tensor, describes the full totally symmetric part
of the interaction to all orders.
We found various ways of rewriting this 4-point coupling in more
symmetric ways, and refer to ref. \cite{Cederwall:2011vy} for the details.

The generalisation to supergravity has not been performed, but should
not present any other difficulties than purely technical, and may be
useful in the search for supersymmetric counterterms. Note that, while
in a component language one must make separate Ans\"atze for the deformed
action and the deformed supersymmetry, here everything is uniformly
encoded in the master equation.

\section{Gauge fixing}
\label{SecGauge}
We will finally briefly mention gauge fixing, which is an important
issue when it comes to quantum calculations and path integrals.

There is a well developed theory of gauge fixing in the BV
framework. One must of course eliminate the antifields as independent
propagating degrees of freedom, and this is achieved by the
introduction of a gauge fermion $\chi$. One then demands that 
\be
\phi^*_I={\delta\chi\over\delta\phi^I}\punkt
\label{GaugeFixFerm}
\ee 
This makes physical quantities independent of gauge choice. Normally,
in a gauge theory, this procedure involves extra non-minimal fields,
the ``antighost'' and Naka\-nishi-Lautrup fields. 

In the pure spinor superfield framework (for maximally supersymmetric
models), we have fields $\Psi$ which effectively contain both fields
and antifields and are self-conjugate under the antibracket. We can
not form a condition like eq. (\ref{GaugeFixFerm}) without a contrived
and unnatural splitting of the field $\Psi$. Therefore it is necessary
to fix the gauge in some other way. 

A standard way to fix gauge in string theory is Siegel gauge 
\cite{Siegel:1988yz}. The gauge fixing condition is 
\be
b\Psi=0\komma
\ee
where $b$ is a ghost field corresponding to the Virasoro constraint.
However, in the pure spinor formalism, no world-sheet or world-line
reparametrisation is \ap\ present --- as we have seen the equations of
motion of the massless fields is an ``indirect'' consequence of
cohomology, and do not follow from ``$p^2=0$'' of some particle
model with reparametrisation symmetry. Such a $b$ operator has to be
constructed as a composite operator if it exists. This was done for
string theory in ref. \cite{Berkovits:2005bt}. The field theory
version of this $b$ operator, relevant for SYM, is
\be
\begin{array}{ll}
b=&-{1\over2}\lalb^{-1}(\lb\g^a D)\*_a
   +{1\over16}\lalb^{-2}(\lb\g^{abc}d\lb)\left(
           N_{ab}\*_c+{1\over24}(D\g_{abc}D)\right)\\
&-{1\over64}\lalb^{-3}(d\lb\g^{abc}d\lb)(\lb\g_a D)N_{bc}\\
&-{1\over1024}\lalb^{-4}(\lb\g^{abe}d\lb)(d\lb\g^{cd}{}_e d\lb)
           N_{ab}N_{cd}\komma
\end{array}
\ee
where $N_{ab}=(\la\g_{ab}{\*\over\*\la})$. The defining property of
the $b$ operator is
\be
\{Q,b\}=\square\punkt
\label{BProperty}
\ee

The whole purpose of gauge fixing is of course to make the kinetic operator
(in this case $Q$) invertible. With this gauge choice, 
the propagator $G$ (``$Q^{-1}$'') is formally
\be
G={b\over\square}\punkt
\ee
So, even if $b$ is a complicated operator, it does precisely what is
needed for gauge fixing: it eliminates almost all the antifields and
implies Lorenz gauge for the gauge connection. By ``almost all'' we
mean that there is a small remainder of the antifield $A^{*a}$,
connected to its on-shell divergencelessness, that gives place for the
antighost, which otherwise is normally introduced by hand. That this
happens follows from the deliberations in ref. \cite{Aisaka:2009yp}.

The consistency of the gauge fixing also relies on the property
$b^2=0$. This identity is quite cumbersome to show --- in string
theory so much so that the full calculation was performed only recently
\cite{Chandia:2010ix,Jusinskas:2013yca}.

In $D=11$ the $b$ operator is quite complicated,
\be
b={1\over2}\eta^{-1}(\lb\g_{ab}\lb)(\la\g^{ab}\g^c D)\*_c+\ldots
\ee
We will not display
it in full detail here, and refer to ref. \cite{Cederwall:2012es}. 

The fact that the $b$ operators, and also other operators carrying
negative ghost number such as the $R^a$ operator of the supergravity
and the physical operators of section \ref{SecBI}, 
have quite complicated expression has been the
source of some activity searching for simpler versions.
See \eg\ refs. 
\cite{Oda:2004bg,Oda:2007ak,Cederwall:2012es,Berkovits:2013pla}.

Once gauge fixing has thus been performed, it is possible to use the
pure spinor superfield formalism for calculation of amplitudes. There
will be further (resolvable) questions about regularisation that we
will completely forgo here, see refs. 
\cite{Berkovits:2006vi,Cederwall:2012es,Bjornsson:2010wm,Bjornsson:2010wu}.
In ref. \cite{Cederwall:2012es}, amplitudes derived from the
supergravity action were shown to be finite up to six loops, in
agreement with refs. \cite{Bjornsson:2010wm,Bjornsson:2010wu} (see the
talk presented by Anna Karlsson, ref. \cite{Karlsson2013}).

It might be expected that gauge fixing in models with less than
maximal supersymmetry can be performed in a way which is more along
the standard lines of the BV formalism, \ie, with a gauge fixing
fermion, since then fields and antifields are naturally separated in
different pure spinor superfields. This remains to be investigated.

\section{Discussion}
\label{SecDisc}
We have given a brief overview of the pure spinor superfield
formalism, and how it leads to off-shell superfield actions for
maximally supersymmetric models. The main focus has been on $D=10$ SYM
and $D=11$ SG, but also other models have been mentioned. Some of the
more technically intricate parts of the formalism have been left out,
but we hope that the general message is clear: this is a solution to
the problem of going off-shell with maximal supersymmetry.

We have repeatedly pointed out the simplicity of the resulting
actions. Indeed, the many terms in a supersymmetric component action
generically reduce to some quite simple expression, which is of lower
order in fields than the component interactions. In a couple of cases,
we even get polynomial expressions where the component ones are
non-polynomial. This is of course an advantage when it comes to
quantum calculations: the number of vertices is very limited. The
other advantage for amplitude calculations is that the presence of an
action (as opposed to a first-quantised formalism) directly yields the
form of the vertices consistent with all symmetries.

The formulation of supergravity has some drawbacks, though. Since only
part of the supervielbein is used, the geometric structure of the
theory is obscured. Background invariance is not manifest, since some
background is needed in order even to define the BRST operator. In
this sense, the behaviour is similar to closed string field theory
\cite{Siegel:1988yz}. 
It is not clear whether geometry, or some aspects of it can be
regained without losing the obvious advantages of the pure spinor
formalism. This means also that solutions beyond the linearised level
around some background are difficult to find, as is \eg\ the dynamics
of extended abjects and their coupling to supergravity.

We believe that there is something to learn from the application of
pure spinor techniques to theories with less supersymmetry. This is
however a largely unexplored subject.

Finally, we would be very interested in extending the formalism to
other structure groups. The type of models we primarily have in mind
are models with ``manifest U-duality'', formulated as gauge theories
within the framework of generalised geometry. Some supermultiplets are
already known in connection with U-duality
\cite{Coimbra:2011ky,Coimbra:2012af,Cederwall:2013naa,Cederwall:2013oaa}, 
and it would be very interesting to continue to a
superfield formalism and maybe a (generalisation of the) pure spinor version.
A manifest control over both supersymmetry and U-duality would be the
ideal situation for examining the ultraviolet properties of maximal
supergravity.

%

%
%
%
%
%
%
%

%
%

%
%



\end{document}